\documentclass[11pt]{article}
\usepackage[letterpaper,margin=1in]{geometry}
\usepackage[T1]{fontenc}
\usepackage[utf8]{inputenc}
\usepackage{amsmath,amssymb,amsfonts}
\usepackage{graphicx}
\usepackage{cite}
\usepackage{microtype}
\usepackage{hyperref}
\hypersetup{colorlinks=true,linkcolor=blue,citecolor=blue,urlcolor=blue}

\title{Stochastic dynamics and synchronization in motif-based neuronal networks}
\author{%
Gurpreet Jagdev$^{1,2}$ \quad Yifei Lu$^{3}$ \quad Richard Bertram$^{4,5,6}$ \quad Na Yu$^{1,2,*}$\\[0.7em]
\small $^{1}$Department of Mathematics, Toronto Metropolitan University, Toronto, Ontario, Canada\\
\small $^{2}$Institute for Biomedical Engineering, Science and Technology (iBEST), Toronto Metropolitan\\
\small University and St. Michael's Hospital, Unity Health Toronto, Toronto, Canada\\
\small $^{3}$Department of Mathematics, University of Toronto, Toronto, Ontario, Canada\\
\small $^{4}$Department of Mathematics, Florida State University, Tallahassee, Florida, USA\\
\small $^{5}$Program in Neuroscience, Florida State University, Tallahassee, Florida, USA\\
\small $^{6}$Institute of Molecular Biophysics, Florida State University, Tallahassee, Florida, USA\\
\small $^{*}$Corresponding author: nayu@torontomu.ca
}
\date{}

\begin{document}
\maketitle
\begin{abstract}
Neuronal networks exhibit complex dynamics shaped by connectivity and stochastic input. Empirical studies show that neuronal networks contain recurring subgraphs, or motifs, but the collective influence of different motif types after embedding in large stochastic networks remains less well understood. We construct a spiking network composed of six representative structural classes and examine how intrinsic noise, coupling strength, inter-motif connectivity, network size, and neuronal heterogeneity shape synchronization. Both motif- and network-level coherence peak at intermediate noise intensities, consistent with coherence resonance. Bidirectionally coupled pairs (M2) and the type-2 recurrent feed-forward loop (M3c) consistently exhibit high motif-level coherence. Rewiring these motifs while preserving local synapse number and strength produces some of the largest reductions in network coherence, showing that connection arrangement contributes beyond strong local coupling alone. M2 and M3c also exhibit frequent spike doublets, linking short inter-spike intervals with elevated coherence. Relative to a degree- and weight-matched random control, the motif-structured network reaches greater coherence at weaker noise. Increasing network size enhances coherence until the response begins to saturate, whereas applied-current heterogeneity lowers peak coherence and shifts the optimum toward stronger noise without erasing the relative differences among motif classes. These results show that local connection arrangement remains dynamically consequential after embedding and can shape noise-driven coherence at the population level.
\end{abstract}

\vspace{0.5em}

\section{Introduction}
\label{sec:introduction}
Spike-timing synchronization is a collective property of neuronal networks and has been associated with distributed information processing in sensory and cognitive systems \cite{varela2001,fell2011}. Its emergence depends on neuronal excitability, synaptic interactions, network structure, and the stochastic fluctuations that are unavoidable in biological neural systems \cite{rosenblum2001}. Noise can disrupt spike timing, but in excitable systems it can also organize activity: at an intermediate intensity, the response may become maximally coherent, a phenomenon known as coherence resonance (CR). The expression of such noise-induced order in a network therefore depends not only on the properties of individual neurons, but also on the paths through which activity propagates.

Neuronal connectivity has structure at several scales. At the local scale, neural circuits contain recurring subgraphs, or network motifs, that occur more often than expected under suitable randomized controls \cite{milo2002network,reigl2004search,song2005highly}. These include bidirectionally coupled pairs, feed-forward loops (FFLs), recurrent FFLs (RFFLs), and larger multi-pathway arrangements. Motifs have been interpreted as candidate computational building blocks, and isolated-motif studies show that their dynamics can depend strongly on internal connection arrangement \cite{mittal2024,jagdev2024,jagdev2024bi,krauss2019,calim2021}. Feedback, for example, can alter stochastic-resonance-like responses in small neuronal circuits \cite{bonsel2022}. Much less is known about whether these motif-specific dynamical differences remain visible after motifs are embedded in a larger network and receive shared background input from many other neurons.

Existing studies of motifs in larger stochastic networks leave this question only partly resolved. Li et al. \cite{li2024} examined stochastic resonance in small-world networks with higher-order motif interactions, but focused on selected FFL outputs rather than comparing several motif classes under common global conditions. More broadly, studies of small circuits often isolate motifs, whereas studies of large neuronal networks typically vary global connectivity without controlling the arrangement of strong local connections. It therefore remains unclear how local motif organization contributes to population-level coherence beyond connection number and mean coupling strength, and how robust such contributions are to changes in network size and neuronal heterogeneity.

We address this problem with a sparse motif-structured network of regular-spiking Izhikevich neurons \cite{IZ}. The network contains a single-neuron baseline and five connected motif classes embedded in a background of weaker Erd\H{o}s--R\'enyi inter-motif connections. We compare this architecture with a random control matched in mean in-degree and synaptic strength, and examine coherence as a function of noise intensity, inter-motif connectivity, the balance of intra- and inter-motif coupling, network size, and applied-current heterogeneity. We also rewire individual motif classes while preserving the number and strength of their local synapses and analyze inter-spike-interval statistics to relate motif structure to spike timing. Across these analyses, the network exhibits CR, but the magnitude of the response depends on local arrangement. M2 and M3c are consistently among the most coherent motif classes, rewiring either produces some of the largest reductions in network coherence, and both generate frequent spike doublets. The motif-structured network also reaches greater coherence at weaker noise than its random control. Together, these comparisons test whether connection arrangement remains dynamically consequential after motifs are embedded in a large stochastic network.

\section{Mathematical model and methods}
\subsection{Network structure}
\label{sec: motif structure}
\begin{figure}[t]
    \centering
    \includegraphics[width=1\linewidth]{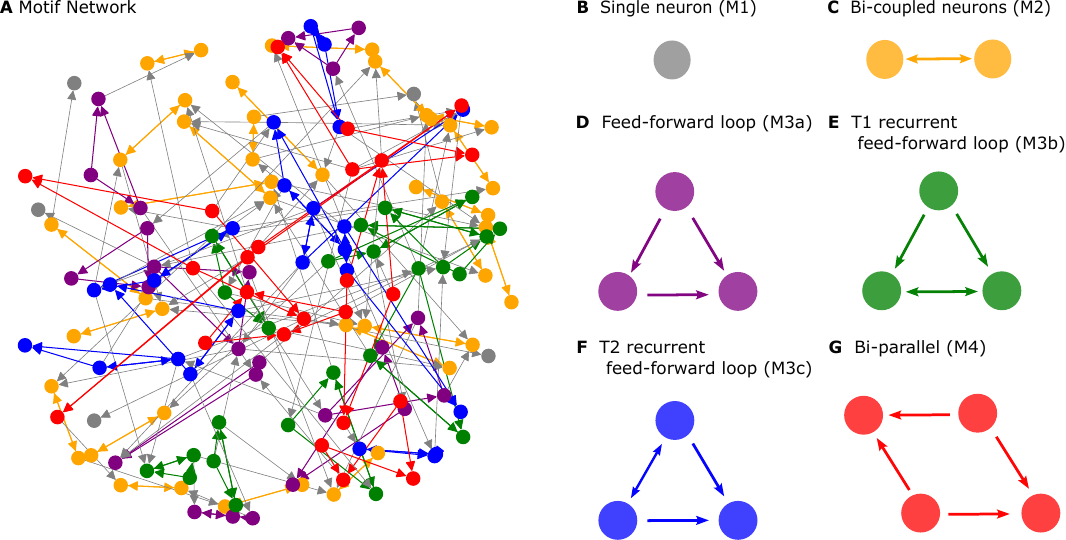}
    \caption{Motif network schematic. (A) Network of 150 neurons with inter-motif connections shown in grey. (B) Motif M1: single neuron (grey). (C) Motif M2: bi-coupled pair (yellow). (D) Motif M3a: feed-forward loop (FFL; purple). (E) Motif M3b: type-1 recurrent FFL (RFFL; green). (F) Motif M3c: type-2 RFFL (blue). (G) Motif M4: bi-parallel motif (red).}

    \label{fig:motif_net}
\end{figure}

The motif network comprises motifs connected internally by strong intra-motif synapses and externally by weaker (random) inter-motif synapses; a 150-neuron schematic is shown in Fig.~\ref{fig:motif_net}A. The selected motifs (Figs.~\ref{fig:motif_net}B–G) are over-represented in neuronal networks \cite{milo2002network,reigl2004search, song2005highly}. Motif M1 (Fig.~\ref{fig:motif_net}B) is a single neuron. M2 (Fig.~\ref{fig:motif_net}C) is a pair of bidirectionally coupled neurons. Motif M3a (Fig.~\ref{fig:motif_net}D) is a three-neuron feed-forward loop (FFL) and motifs M3b,c (Figs.~\ref{fig:motif_net}E and F, respectively) are three-neuron recurrent FFLs (RFFLs). M3b, a type-1 (T1) RFFL, features a recurrent (i.e. bidirectional) connection from the intermediate to output neuron, whereas M3c, a type-2 (T2) RFFL, features a recurrent connection from the input to intermediate neuron. Lastly, M4 (Fig.~\ref{fig:motif_net}G) is a four-neuron bi-parallel motif.

In terms of absolute frequency, \textit{simple} motifs such as M1 and M2 occur more often than structurally complex motifs such as M3a–c and M4. Motifs M3a–c and M4, however, are overrepresented relative to random networks, indicating non-random structural organization and potential functional significance \cite{milo2002network, song2005highly}. Thus, the distribution of motifs is chosen to reflect functional importance rather than absolute frequency. Neurons belonging to three-neuron motifs account for 45\% of the network, with M3a-c each comprising 15\%. Single neurons (M1) are minimally represented at 9\% of the network. Between motifs M2 and M4, M2 neurons are assigned greater weight (30\%) than M4 neurons (16\%) due to probabilistic considerations. For a network of 1000 neurons, this corresponds to 90 M1 motifs (90 neurons), 150 M2 motifs (300 neurons), 50 of each M3a, b, and c motifs (450 neurons total), and 40 M4 motifs (160 neurons).

Neurons belonging to distinct motifs (of the same or different kind, e.g., grey lines in Fig.~\ref{fig:motif_net}) are connected via inter-motif synapses according to an Erd\H{o}s--R\'enyi model. That is, neurons $i$ and $j$ are connected with probability $p$ only if $|i - j| > 4$, this ensures that inter-motif edges connect neurons of distinct motifs---the largest of which, M4, contains four neurons.

\subsection{Mathematical model}

To model individual neuron dynamics, we adapt the Izhikevich (IZ) model \cite{IZ}. Each neuron is governed by the system of stochastic differential equations:
\begin{align}  
\frac{\mathrm{d}v_i}{\mathrm{d}t} &= 0.04v_i^2 + 5v_i + 140 - u_i + I + I_{\text{syn}}^{(i)} + D\eta_i(t), \label{v} \\
\frac{\mathrm{d}u_i}{\mathrm{d}t} &= a \big( bv_i - u_i \big), \label{u} 
\end{align}
subject to the after-spike resetting rule,
\begin{equation} 
\text{if } v_i \geq 30 \, \text{mV, then } 
\label{reset}
\begin{cases}
v_i = c \\
u_i = u_i + d
\end{cases}, \quad i \in \{1, 2, \ldots, N\}.
\end{equation}
$N=1000$ is the network size and time, $t$, is measured in milliseconds (ms). $v_i$ (mV) represents the neuron’s membrane potential, and $u_i$ (mV) is a recovery variable that models a slow negative feedback process regulating $v_i$, corresponding to the combined effects of potassium activation and sodium inactivation currents. $I$ (mA) is an applied current and control parameter. The total (incoming) presynaptic current consists of intra- and inter-motif components: $I_{\text{syn}}^{(i)}(t) = I_{\text{syn,intra}}^{(i)}(t) + I_{\text{syn,inter}}^{(i)}(t)$. Synapses are modelled as pulse-coupled connections—well suited to approximate chemical synapses in reduced models \cite{IZ1999class}—where a presynaptic spike at time $\tau^f$ induces a postsynaptic impulse. That is,
\begin{equation}
\label{syn_pulse}
I_{\text{syn}}^{(i)}(t) = |\mathcal{P}_i|^{-1} \sum_{j \in \mathcal{P}_i} W_{i,j} \sum_{\tau^f_j \in \mathcal{F}_j} \delta(t - \tau^f_j),
\end{equation}
where $\mathcal{P}_i$ is the set of presynaptic connections for neuron $i$, $\mathcal{F}_j$ is the set of presynaptic spike times, and $W_{i,j} > 0$ is the synaptic weight. We distinguish $W_{i,j} = W_{\text{intra}}$ for intra-motif and $W_{i,j} = W_{\text{inter}}$ for inter-motif connections.  $D\eta_i(t)$ is an intrinsic noise stimulus modelled as a Wiener process $\eta_i(t)$ with zero mean, unit variance, and intensity $D$. The parameters $a$, $b$, $c$, and $d$ govern the recovery and reset dynamics: $a$ controls the recovery timescale of $u_i$, $b$ determines the sensitivity of $u_i$ to subthreshold oscillations in $v_i$, $c$ sets the post-spike reset value of $v_i$, and $d$ sets the post-spike increment of $u_i$.

We consider the \textit{regular spiking} regime of the IZ model, which mimics the most prominent firing patterns in cortical cells \cite{IZ}. This regime is defined by a deep voltage reset ($c = -65$) and a large recovery increment ($d = 8$). For $b > a$, the resting state loses stability via a Hopf bifurcation (HB) of the resting state. Biophysically plausible parameters are $a = 0.02$ and $b = 0.2$ \cite{IZ}.

\subsection{Bifurcation analysis of the homogeneous deterministic model}
\label{sec:bifurcation analysis}
In the deterministic and uncoupled regime ($D,p=0$), the equilibria of Eqs.~\eqref{v}–\eqref{u} satisfy $\dot{v}=\dot{u}=0$, yielding
\begin{equation*}
    (v^*,u^*) = \left(12.5\left[b - 5 \pm \sqrt{(5-b)^2 - 0.16(140 + I)}\right], \, b v^*\right).
\end{equation*}
These equilibria change stability as the applied current $I$ traverses two critical points,
\begin{equation*}
    I = 6.25\left[(5-b)^2 - (b-a)^2\right] - 140
\ \text{  and  } \
I = 6.25(5-b)^2 - 140,
\end{equation*}
corresponding to Hopf and fold bifurcations, respectively. Linearizing the system about $(v^*,u^*)$ gives the characteristic polynomial of the Jacobian,
\begin{equation*}
\lambda^2 - (0.08v^* + 5 - a)\lambda - a(0.08v^* + 5) + ab = 0,
\end{equation*}
whose eigenvalues determine local stability. For the regular-spiking parameters $a=0.02$ and $b=0.2$, the system exhibits critical points at $I \approx 3.8$ and $I = 4$. When $I < 3.8$, the neuron has two equilibria---a stable node (resting state) and a saddle node (rheobase)---and small perturbations decay back to rest, defining the excitable regime. For $3.8 < I < 4$, both equilibria become unstable, producing sustained oscillations under the reset rule Eq.~\eqref{reset}, while for $I > 4$, the equilibria vanish and the system undergoes self-sustained tonic firing. Thus, $I < 3.8$ corresponds to a quiescent regime in which noise-driven excitation can be isolated and studied. A complete derivation of these bifurcation conditions is provided in \textit{Appendix~\ref{appendix:bifurcation}}.

\subsection{Random network}
\label{sec: random network}
To assess motif-network performance relative to a control, we construct a random network that preserves both the in-degree distribution and the mean synaptic input of the motif network. As inter-motif connections follow an Erd\H{o}s--R\'enyi scheme, the motif network exhibits a binomial in-degree distribution, with intra-motif connections contributing negligibly. The random network is therefore defined by the degree distribution $\text{Binomial}(\widehat{N},\ \overline{p})$, where $\widehat{N} = N - 9$ accounts for the distance constraint $|i - j| > 4$, and $\overline{p} = (N\widehat{N})^{-1} \sum_{i=1}^N |\mathcal{P}_i|$ matches the empirical mean in-degree of the motif network. The corresponding mean synaptic weight is
$
\overline{W} = \sum_{i=1}^N |\mathcal{P}_i|^{-1} \big( W_\text{intra} |\mathcal{P}_{\text{intra},i}| + W_\text{inter} |\mathcal{P}_{\text{inter},i}| \big),
$
where $\mathcal{P}_{\text{intra},i}$ and $\mathcal{P}_{\text{inter},i}$ denote the intra- and inter-motif presynaptic sets of neuron $i$. For parameters $W_\text{intra} = 30$, $W_\text{inter} = 15$, and $p = 0.05$ in the motif network, the random control has $\overline{p} \approx 0.0507$ and $\overline{W} = 15.3$. 

\subsection{Methods}
\label{sec: methods}
Simulations are performed using the Brian2 package in Python \cite{stimberg2019}. Eqs. \ref{v}–\ref{reset} are integrated via the Euler–Maruyama method with time-step $\text{dt} = 0.5\,\text{ms}$ over a total duration of $T = 2000\,\text{ms}$, discarding an initial transient of $t_0 = 100\,\text{ms}$. Initial conditions are assigned per neuron and trial as $v_i(0) = cQ$, $Q \sim \text{Uniform}[0,1]$, and $u_i(0) = b v_i(0)$, following \cite{IZ}. Simulation data are exported and analyzed in MATLAB. Reported results represent averages over 50 independent trials with randomized initial conditions and inter-motif connectivity. Figures are produced using Python or MATLAB and finalized in the graphics editor Inkscape. 

\section{Results}
\label{sec:results}
\subsection{Noise-induced dynamics}
\begin{figure}[t]
    \centering
    \includegraphics[width=1\linewidth]{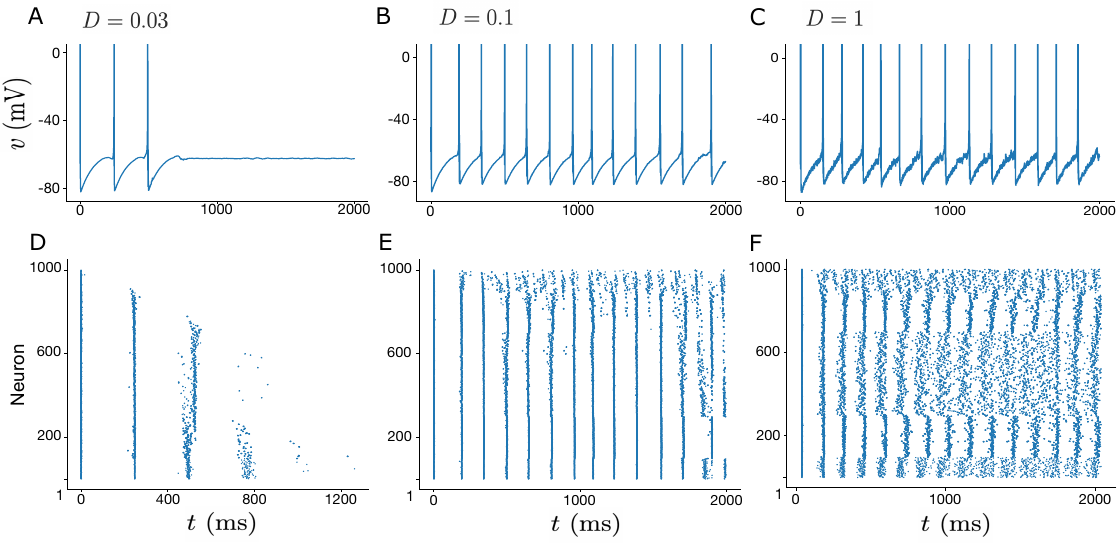}
   \caption{Timeseries and raster plots of membrane potential, $v$, under different noise intensities, $D$. (A)–(C) $v$-timeseries for neuron 500 at $D = 0.03$, $0.1$, and $1$, respectively. (D)–(F) Corresponding spike raster plots for $D = 0.03$, $0.1$, and $1$. Parameters: $I = 3.75$, $W_\text{intra} = 30$, $W_\text{inter} = 15$, and $p = 0.05$.}
    \label{fig:timeseries}
\end{figure}

We examine the network dynamics in the excitable regime ($I < 3.8$), near the onset of a subcritical HB. As shown in Figs.~\ref{fig:timeseries}A–C, the network remains quiescent in the absence of noise and exhibits noise-driven activity as the intensity $D$ increases. For weak noise ($D = 0.03$; Fig.~\ref{fig:timeseries}A), neurons display transient spiking with low-amplitude subthreshold oscillations. Increasing the noise intensity ($D = 0.1$; Fig.~\ref{fig:timeseries}B) induces tonic spiking characterized by regular inter-spike intervals (ISIs), whereas strong noise ($D = 1$; Fig.~\ref{fig:timeseries}C) leads to irregular firing patterns, indicating loss of coherence due to excessive stochastic forcing.  

The corresponding raster plots (Figs.~\ref{fig:timeseries}D–F) confirm these observations at the population level. At moderate noise ($D = 0.1$), spikes occur with consistent timing across neurons, whereas at high noise ($D = 1$) spike events become temporally dispersed. Under strong noise, motif-specific behaviour emerges (Fig.~\ref{fig:timeseries}F): neurons within motifs M2 (indices 91–301) and M3c (indices 540–689) exhibit higher robustness to noise, maintaining coherent spiking, while the neurons in other motifs exhibit irregular, asynchronous dynamics.

\subsection{Noise-induced synchrony}
In this study, \textit{synchrony} and \textit{coherence} refer to the similarity of spike timing across neurons; higher synchrony therefore indicates more consistent spike trains across an ensemble. To quantify synchrony we apply the signal-to-noise ratio (SNR) measure:
\begin{equation}
\label{SNR}
\text{SNR} = \frac{1}{N}\sum_{i=1}^N \frac{P_{\text{signal}, i}}{P_{\text{noise}, i}}
= \frac{1}{N}\sum_{i=1}^N \sqrt{\frac{\left\lvert \frac{1}{M} \sum_{m=1}^M X_i^2[m] \right\rvert}{D^2}},
\end{equation}
where $X_i[m] = \text{DFT}\{x_i\}$ is the discrete Fourier transform of the binary spike train $x_i$, $m \in \{1,2,\ldots,M\}$ is the frequency index, and $M$ is the number of sampled points. $P_{\text{signal}, i}$ and $P_{\text{noise}, i}$ denote the signal and noise power, respectively. Since $D\eta_i(t)$ is a white-noise process, $P_{\text{noise}, i} = D$ for all $i$. The SNR thus serves as an input–output measure of spike-timing coherence, with larger values indicating stronger synchrony \cite{czanner2015measuring}.

\begin{figure}[t]
    \centering
    \includegraphics[width=\linewidth]{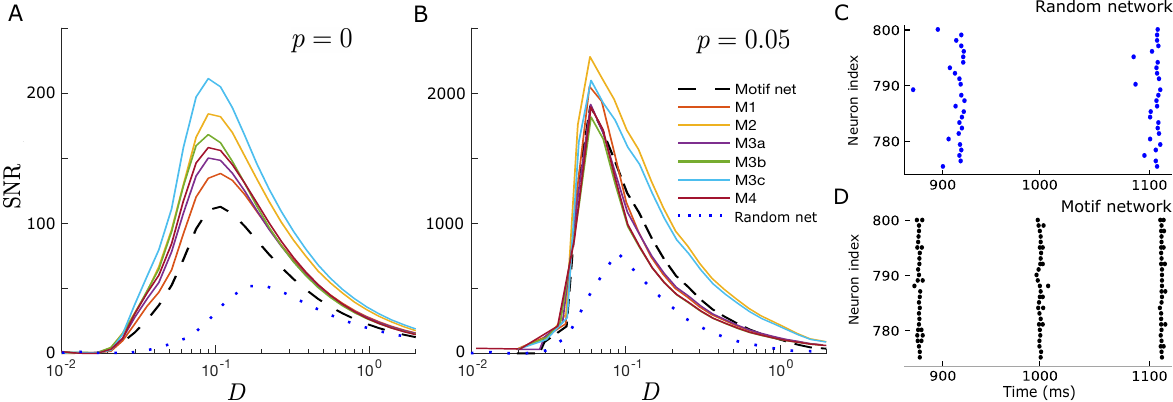}
   \caption{Noise-induced synchronization per motif. (A)–(B) SNR as a function of $D$ for motifs M1 (orange), M2 (yellow), M3a (purple), M3b (green), M3c (blue), and M4 (burgundy), as well as the motif network (dashed line) and a random network (dotted line), for inter-motif/random connection probabilities (A) $p = 0$ and (B) $p = 0.05$. For the motif network: $W_\text{intra} = 30$, $W_\text{inter} = 15$, and $I = 3.75$; for the random network: $\overline{p} = 0.0507$ and $\overline{W} = 15.3$. (C)–(D) Raster plots for the random (C) and motif (D) networks with $D = 0.07$.}

    \label{fig:snr_clusters}
\end{figure}

The dependence of the SNR on $D$ is shown in Figs.~\ref{fig:snr_clusters}A and B, where the dashed black curves represent the network-level SNR for unconnected ($p=0$; Fig.~\ref{fig:snr_clusters}A) and connected ($p=0.05$; Fig.~\ref{fig:snr_clusters}B) networks. In both cases, the SNR varies non-monotonically with $D$: it increases for weak noise, peaks at an intermediate, optimal intensity $D^*$, and declines for stronger noise. This behaviour is indicative of coherence resonance (CR), where intermediate noise maximizes spike-timing regularity in excitable systems \cite{gang1993stochastic}. The inclusion of inter-motif coupling reduces the optimal noise intensity from $D^* \approx 0.1$ when $p=0$ to $D^* \approx 0.06$ when $p=0.05$ while also increasing the peak SNR from $\text{SNR}^* \approx 110$ when $p=0$ to $\text{SNR}^* \approx 1800$ when $p=0.05$.

Per motif synchrony is evaluated by computing the SNR over all neurons belonging to a certain motif type. All else equal, each motif reaches maximal coherence at similar noise levels, even in the uncoupled case. However, the degree of synchrony varies. When $p=0$ (Fig.~\ref{fig:snr_clusters}A), motif M3c (blue) achieves the highest SNR, followed by M2 (yellow). M1 (orange), lacking internal coupling, yields the lowest SNR, while M3a (purple), M3b (green), and M4 (burgundy) display intermediate coherence, with M3b marginally exceeding M3a and M4. With the inclusion of inter-motif connections ($p=0.05$; Fig.~\ref{fig:snr_clusters}B), this ordering remains consistent, with M3c slightly surpassing M2.

In comparison to the random control network, the motif network exhibits higher SNR across all $D$. The motif network (dashed black curve) outperforms the random control (dotted blue curve) for both $p,\overline{p}=0$ (Fig.~\ref{fig:snr_clusters}A) and $p=0.05$, $\overline{p}=0.0507$ (Fig.~\ref{fig:snr_clusters}B). At $\overline{p}=0.0507$, the random network attains $\text{SNR}^* \approx 800$ at $D^* \approx 0.1$, whereas the motif network reaches $\text{SNR}^* \approx 1800$ at $D^* \approx 0.05$. Structured motif connectivity therefore enhances noise-driven synchrony relative to a degree- and weight-matched random architecture. This disparity is further evidenced by the raster plots in Figs.~\ref{fig:snr_clusters}C–D): the motif network (Fig.~\ref{fig:snr_clusters}D) exhibits higher firing rates, doublet bursting (Sec.~\ref{sec: doublet burst}), and greater temporal coherence across neurons, whereas the random network (Fig.~\ref{fig:snr_clusters}C) shows sparser and more irregular spiking activity.

\subsection{Effect of inter-motif connectivity}
\label{sec: inter-motif probability}
\begin{figure}[t]
    \centering
    \includegraphics[width=\linewidth]{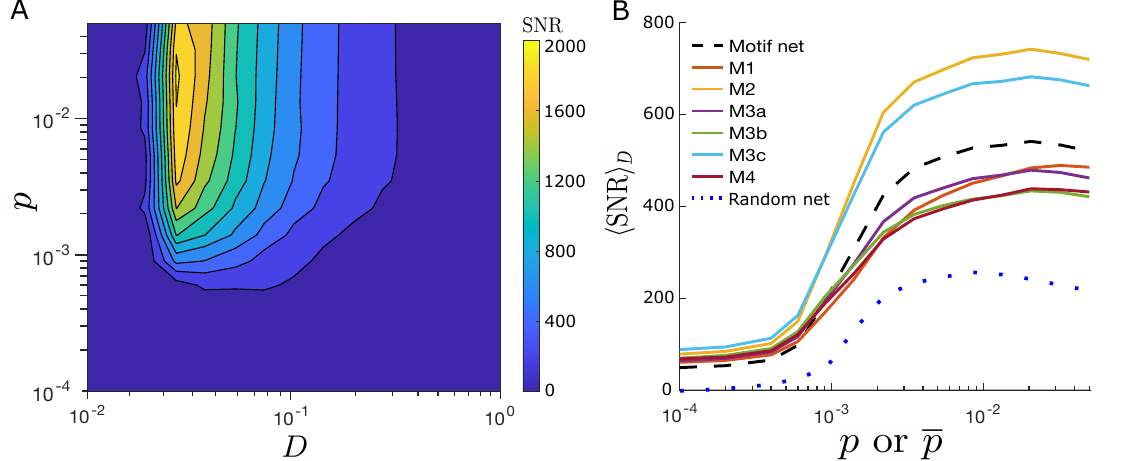}
   \caption{Effect of inter-motif connectivity. (A) Heat map of SNR as a function of $p$ (or $\overline{p}$) and $D$, with warmer colors indicating higher SNR. (B) Noise-averaged SNR, $\langle \mathrm{SNR} \rangle_D$, versus $\overline{p}$ for the random network (dotted line) and $p$ for motifs M1 (orange), M2 (yellow), M3a (purple), M3b (green), M3c (blue), M4 (burgundy), and the motif network (dashed line). Parameters: motif network---$W_\text{intra} = 30$, $W_\text{inter} = 15$; random network---$\overline{W} = 15.3$.}
    \label{fig:connectionprob}
\end{figure}

The influence of inter-motif connectivity on noise-induced synchrony is evaluated by computing the SNR (Eq.~\ref{SNR}) across varying $D$ and $p$. The results are presented as a heat map in Fig.~\ref{fig:connectionprob}A, where warmer colours denote higher SNR. In addition, the noise-averaged SNR, $\langle \text{SNR} \rangle_D = |\mathcal{D}|^{-1}\sum_{D' \in \mathcal{D}} \text{SNR}(D')$, where $\mathcal{D}$ is the set of sampled noise intensities, is plotted in Fig.~\ref{fig:connectionprob}B as a function of $p$ for the motif network, random control, and individual motif types
For weak inter-motif coupling ($p < 0.0006$), the noise-averaged SNR $\langle \text{SNR} \rangle_D$ remains stable at $\approx 100$. In the interval $0.0006 \leq p \leq 0.002$, $\langle \text{SNR} \rangle_D$ increases sharply, marking a transition in network coherence. This region corresponds to a heightened sensitivity to $p$, where small increments in connectivity yield large gains in synchrony (Fig.~\ref{fig:connectionprob}A–B). For $p > 0.002$, both the heat map and $\langle \text{SNR} \rangle_D$ curves approach saturation, indicating reduced sensitivity to further increases in $p$. The network-level SNR attains a maximum near $p \approx 0.02$ ($\text{SNR} \approx 2000$) and remains nearly constant for larger $p$, with a slight decline to $\approx 1800$ at $p=0.05$. Across this region ($p \geq 0.0006$), the optimal noise intensity remains fixed at $D^* \approx 0.04$.

Although $\langle \text{SNR} \rangle_D$ increases with the increment of $p$ for all motifs, the rate and magnitude of this increase differ across motif types (Fig.~\ref{fig:connectionprob}B). M2 and M3c exhibit the largest coherence gain, whereas the feed-forward loop motif M3a shows a weaker response. M4, despite containing the most neurons, fails to reach high synchrony, indicating that motif size (number of neurons or connections) alone does not determine sensitivity to inter-motif coupling. The three feed-forward loop motifs (M3a–c), although identical in neuron and synapse count, display distinct responses to $p$, demonstrating that internal synaptic organization governs their coherence non-trivially. As $p$ increases, motif responses diverge: for $p < 0.0006$, $\langle \text{SNR} \rangle_D$ values are nearly identical, while for $p > 0.002$, the curves separate, revealing distinct coherence levels. The relative ordering, however, remains consistent with Fig.~\ref{fig:snr_clusters}B: M2 exhibits the highest $\langle \text{SNR} \rangle_D$, followed by M3c, while M3a, M1, M3b, and M4 cluster at lower coherence levels, below or closely aligned with the network average.

\subsection{Interplay of intra- and inter-motif coupling strengths}
\label{sec: The interplay of coupling strengths}

Synaptic coupling strengths critically determine the emergence of local and global synchronization by regulating how neurons integrate inputs within and across motifs. We analyze how intra- and inter-motif coupling strengths ($W_{\text{intra}}$ and $W_{\text{inter}}$) shape noise-induced synchrony at both the motif and network levels.

\begin{figure}[t]
    \centering
    \includegraphics[width=\linewidth]{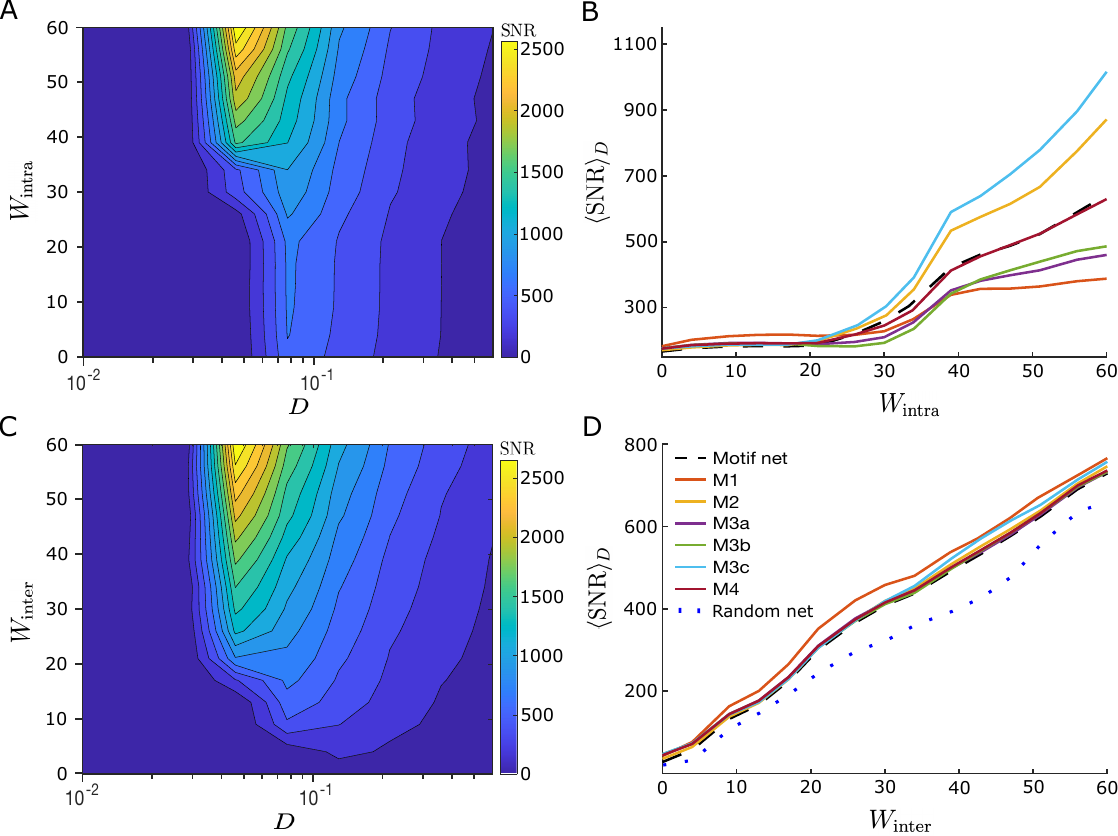}
    \caption{Impact of coupling strengths on synchrony. (A) Heat map of SNR versus $W_\text{intra}$ and $D$. (B) Heat map of SNR versus $W_\text{inter}$ and $D$. (C) $\langle \mathrm{SNR} \rangle_D$ as a function of $W_\text{intra}$ per motif or $\overline{W}$ for the random network (dotted line). (D) $\langle \mathrm{SNR} \rangle_D$ versus $W_\text{inter}$ per motif. Lines: M1 (orange), M2 (yellow), M3a (purple), M3b (green), M3c (blue), M4 (burgundy), and the motif network (dashed). In (A) and (B), $W_\text{inter} = 10$; in (C) and (D), $W_\text{intra} = 10$. Parameters: $I = 3.75$, $p = 0.05$ (motif network), and $\overline{p} = 0.0507$ (random network).}
    \label{fig:coupling_strengths}
\end{figure}

We first vary $W_{\text{intra}}$ while fixing $W_{\text{inter}} = 5$. Fig.~\ref{fig:coupling_strengths}A shows the resulting SNR as a function of $W_{\text{intra}}$ and $D$ in the form of a heat map. Across all $W_{\text{intra}}$, a distinct SNR maximum indicates CR, suggesting that noise-induced synchrony occurs even at minimal coupling. Moreover, the SNR is positively correlated with $W_{\text{intra}}$. For example, the peak SNR rises from $\approx 750$ at $W_{\text{intra}} \approx 20$ to $\approx 2200$ at $W_{\text{intra}} \approx 50$. For $W_{\text{intra}} \leq 35$, the peak remains nearly constant, while a sharp transition to higher coherence occurs between $W_{\text{intra}} = 35$ and $40$, beyond which SNR continues to grow gradually. The optimal noise intensity $D^*$ exhibits a tri-partite structure: for $W_{\text{intra}} \leq 35$, $D^* \approx 0.08$; for $W_{\text{intra}} \geq 40$, $D^* \approx 0.05$; and between $35 \leq W_{\text{intra}} \leq 40$, $D^*$ decreases smoothly between these values. This transition band marks a critical shift in the networks sensitivity to noise.

Conversely, varying $W_{\text{inter}}$ with $W_{\text{intra}} = 5$ (Fig.~\ref{fig:coupling_strengths}C) yields a bi-partite pattern in $D^*$. For $W_{\text{inter}} \leq 20$, $D^*$ decreases steadily from $\approx 0.15$ to $0.05$ with the increment of $W_{\text{inter}}$, beyond which it stabilizes. Unlike $W_{\text{intra}}$, increasing $W_{\text{inter}}$ produces a uniform rise in peak SNR. Furthermore, the comparable SNR ranges in Figs.~\ref{fig:coupling_strengths}A and C may suggest $W_{\text{intra}}$ exerts a disproportionately strong effect on network dynamics: despite representing fewer connections, intra-motif coupling contributes comparably to overall synchrony.

To assess motif-specific effects, Figs.~\ref{fig:coupling_strengths}B and D show $\langle \text{SNR} \rangle_D$ as a function of $W_{\text{intra}}$ and $W_{\text{inter}}$, respectively. In Fig.~\ref{fig:coupling_strengths}D, all motifs exhibit an approximately linear increase with $W_{\text{inter}}$, consistent with the random inter-motif connectivity. Notably, M1 neurons---despite lacking internal coupling---achieve the highest $\langle \text{SNR} \rangle_D$, surpassing all other motifs including the network average. This likely reflects their strong dependence on inter-motif input; however, their higher coherence relative to structured motifs (e.g., M2, M3c) is unexpected. Consistent with earlier results, the random control (dotted blue curve) yields consistently lower $\langle \text{SNR} \rangle_D$ across all $W_{\text{inter}}$ and $\overline{W}$, remaining below both the motif network and individual motifs, further corroborating our finding that structured connectivity promotes stronger noise-driven synchronization relative to completely random analogues.

In Fig.~\ref{fig:coupling_strengths}B, when $W_\text{intra} \leq 20$, all motifs except M1 maintain near-constant $\langle \text{SNR} \rangle_D \approx 50$, while M1 exhibits elevated synchrony ($\approx 100$), consistent with Fig.~\ref{fig:coupling_strengths}D. Between $20 \leq W_\text{intra} \leq 40$, synchrony increases sharply in a motif-dependent manner: M3c and M2 exhibit the largest gains, exceeding the network average, while M3a, M3b, and M1 lag behind. For $40 \leq W_\text{intra} \leq 60$, the relative rankings persist but the growth rate plateaus. Within this range, M3b overtakes M3a and M1, rendering M1 the least coherent motif. 

\begin{figure}[t]
    \centering
    \includegraphics[width=\linewidth]{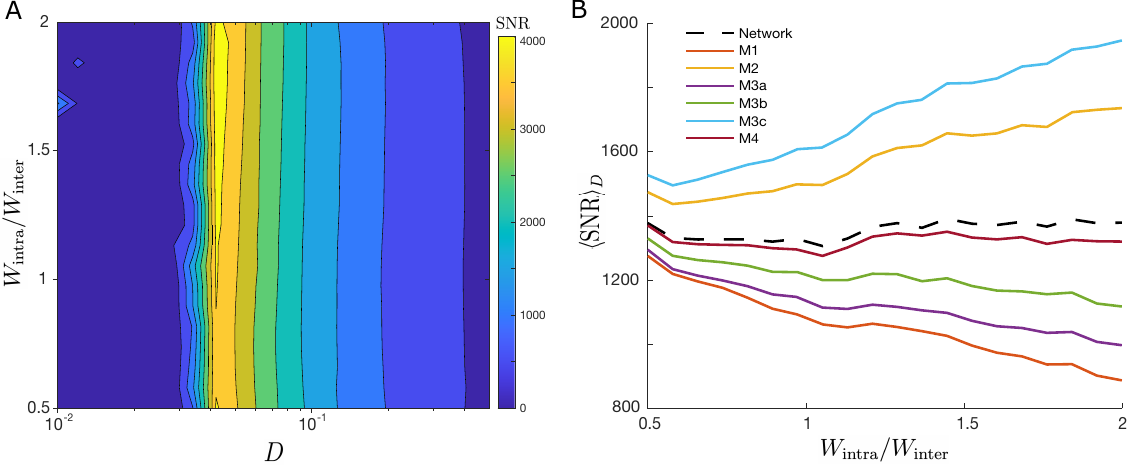}
   \caption{Interplay of inter- and intra-motif coupling strengths. (A) Heat map of SNR versus $W_\text{intra}/W_\text{inter}$ and $D$. (B) $\langle \mathrm{SNR} \rangle_D$ versus $W_\text{intra}/W_\text{inter}$ for motifs M1 (orange), M2 (yellow), M3a (purple), M3b (green), M3c (blue), M4 (burgundy), and the motif network (dashed). Parameters: $I = 3.75$ and $p = 0.05$.}

    \label{fig:synapse_ratio}
\end{figure}

Having examined the effects the intra- and inter-motif coupling strengths in isolation, we now investigate how their relative magnitudes jointly modulate synchrony. That is, we fix total coupling strength at $W_\text{intra} + W_\text{inter} = 100$ and vary the ratio $W_\text{intra}/W_\text{inter}$ from 0.5 to 2. The resulting SNR and $\langle \text{SNR} \rangle_D$ curves are shown in Fig.~\ref{fig:synapse_ratio}. Across all ratios, the network exhibits clear CR with stable optimal noise intensity $D^* \approx 0.05-0.06$, and peak SNR increasing modestly from $\approx 3500$ at $W_\text{intra}/W_\text{inter} \leq 1$ to $\approx 4000$ for $W_\text{intra}/W_\text{inter} \geq 1$ (Fig.~\ref{fig:synapse_ratio}A). Notably, the peak region of the SNR heat map broadens slightly with increasing $W_\text{intra}/W_\text{inter}$, indicating that stronger intra-motif coupling expands the range of noise intensities that support optimal coherence.

Fig.~\ref{fig:synapse_ratio}B illustrates how the intra- to inter-motif balance influences motif-level coherence. All motifs display approximately linear dependence on $W_\text{intra}/W_\text{inter}$ but with distinct slopes. M3c and M2 exhibit the steepest increases, indicating maximal benefit from greater relative intra-motif coupling. M4 shows weak sensitivity to this ratio, remaining slightly below the network average. In contrast, M3a, M3b, and M1 exhibit decreasing synchrony with increasing ratio. While M1’s decline reflects its lack of intra-motif connections, the divergence among FFL motifs is nontrivial: M3c gains substantially from intra-motif coupling, whereas M3a and M3b do not. The relative motif rankings corroborate our former findings, with motifs M3c and M2 consistently outperforming the network average and other motifs, reinforcing their roles as coherence-promoting structures.

\subsection{Effects of motif rewiring on network synchrony}

\begin{figure}[t]
\centering
\includegraphics[width=0.7\linewidth]{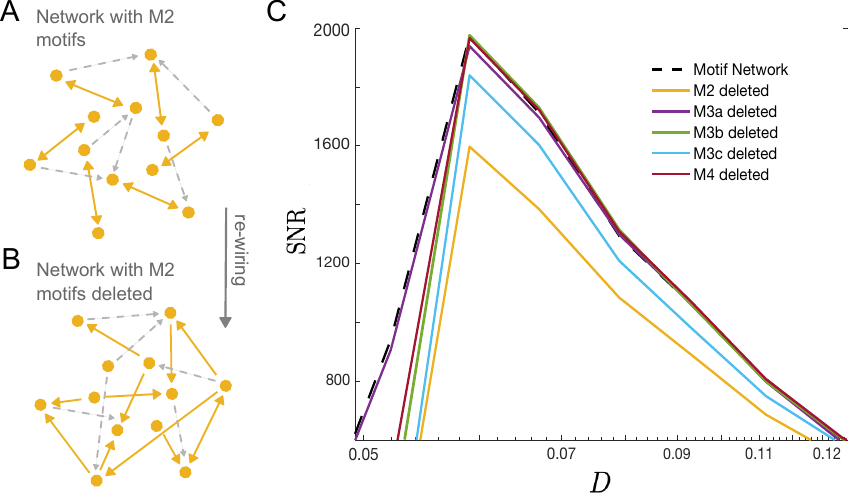}
\caption{Motif rewiring. (A) Sub-network of M2 motifs. (B) Rewired sub-network with M2 motifs deleted. Yellow arrows denote intra-motif connections and dashed grey arrows indicate random inter-motif connections ($p = 0.05$). (C) SNR as a function of $D$ after motif rewiring. The dashed black curve represents the intact motif network; the yellow, purple, green, blue, and burgundy curves correspond to deletion of motifs M2, M3a, M3b, M3c, and M4, respectively. Parameters: $W_\text{intra} = 30$, $W_\text{inter} = 15$, and $p = 0.05$.}
\label{fig:deletion}
\end{figure}

To further assess the contribution of specific motif types to network coherence, we perform a motif-rewiring analysis in which all instances of a given motif are replaced by randomized local sub-networks that preserve the number of synapses and the weight of each connection. This modification removes structured intra-motif connectivity while maintaining strong short-range coupling, yielding a more direct comparison than a globally random network with only matched degree and weight statistics. As an example, Figs.~\ref{fig:deletion}A,B show an M2 sub-network and its randomized counterpart, illustrating the motif-rewiring procedure.

Fig.~\ref{fig:deletion}C shows the SNR of the motif network after systematically rewiring specific motif types. Rewiring M2 or M3c produces a pronounced reduction in SNR across all noise intensities, with both curves lying below the network average, indicating that these motifs play a dominant role in supporting coherent spiking. For instance, the intact network reaches a peak SNR of $\approx 2000$ at $D^*\approx 0.06$, whereas rewiring of M2 reduces the peak to $\approx 1600$, and rewiring of M3a lowers it to $\approx 1800$. In contrast, removal of M3a yields a smaller decrease, as the blue and dashed black curves remain closely aligned near the peak region, while rewiring of M3b or M4 slightly elevates the SNR to values $\approx 2100$, suggesting that these motifs weakly suppress network synchrony. The position of the SNR maximum remains fixed at $D^*\approx 0.06$ for all deletion cases, implying that the optimal noise level for CR is invariant to local structural perturbations. The overall shape of the SNR curves is preserved, indicating that motif rewiring modulates the amplitude of CR rather than its form. For strong noise ($D > D^*$), all rewiring curves except those of M2 and M3c converge with the network average, indicating that stochastic forcing dominates over structural effects. For weak noise ($D < D^*$), however, rewiring of any motif other than M3a slightly decreases the SNR relative to the intact network. Overall, rewiring M2 yields the largest reduction in coherence, followed by M3c, corroborating earlier results that these motifs most effectively promote noise-induced synchronization.

\subsection{Doublet formation and network synchrony}
\label{sec: doublet burst}
\begin{figure}[t]
    \centering
    \includegraphics[width=1\linewidth]{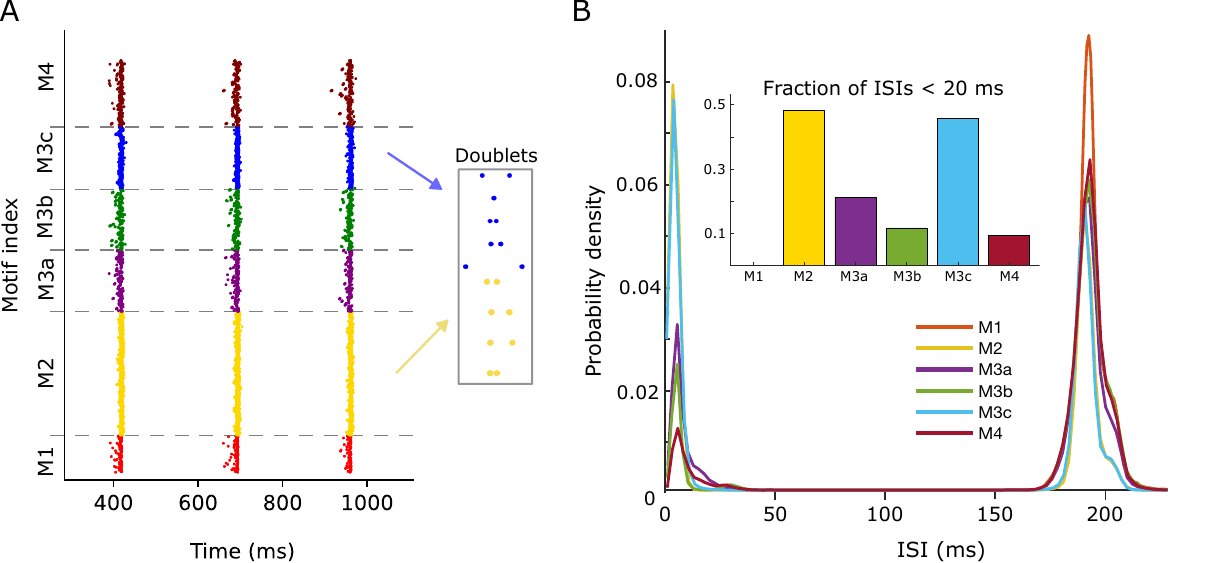}
 \caption{Doublet formation and ISI distribution per motif. (A) Raster plot showing an example of doublet bursting for motifs M2 and M3c. (B) Empirical probability density functions of the ISI distribution per motif. The inset shows the fraction of ISIs shorter than 20~ms for each motif. Colors: M1 (red), M2 (yellow), M3a (purple), M3b (green), M3c (blue), and M4 (burgundy). Parameters: $p = 0.05$, $W_\text{intra} = 30$, $W_\text{inter} = 15$, and $I = 3.75$.}
    \label{fig:doublet}
\end{figure}

To examine the structural origin of enhanced coherence in specific motif types, particularly M3c and M2, we analyze the spiking statistics of individual motifs and identify a form of noise-induced bursting characterized by spike doublets. Fig.~\ref{fig:doublet}A displays a sample raster plot in which neurons within M2 and M3c motifs exhibit doublet bursting, producing pairs of closely spaced spikes separated by short inter-spike intervals (e.g., $<20$~ms). The inset of Fig.~\ref{fig:doublet}A provides a magnified view of the doublet events, illustrating the bursting dynamics in M2 and M3c.

Fig.~\ref{fig:doublet}B shows the ISI probability density for each motif together with the fraction of short ISIs ($<20$~ms; inset). All motifs except M1 display noise-induced doublet bursting to varying degrees. Motifs M2 and M3c exhibit the highest proportion of short ISIs ($\approx0.485$ and $\approx0.47$, respectively), indicating frequent doublet formation. In contrast, M4 displays only a small fraction of short ISIs, and M1 shows none, consistent with its lack of internal coupling and isolated spiking with ISIs centered near $\approx190$~ms. Interestingly, M3a---which lacks recurrence---produces a greater proportion of short ISIs than M3b, which contains a recurrent connection, underscoring that recurrence alone is insufficient to generate doublet bursting. The relative differences among motifs indicate that the placement of recurrent connections, rather than their mere presence, determines the likelihood of doublet formation. Furthermore, we hypothesize that the fraction of short ISIs may contribute to a motif’s propensity for noise-induced synchrony and high SNR, but in a nontrivial manner determined by its specific internal connection topology.

\begin{figure}[t]
\centering
\includegraphics[width=0.9\linewidth]{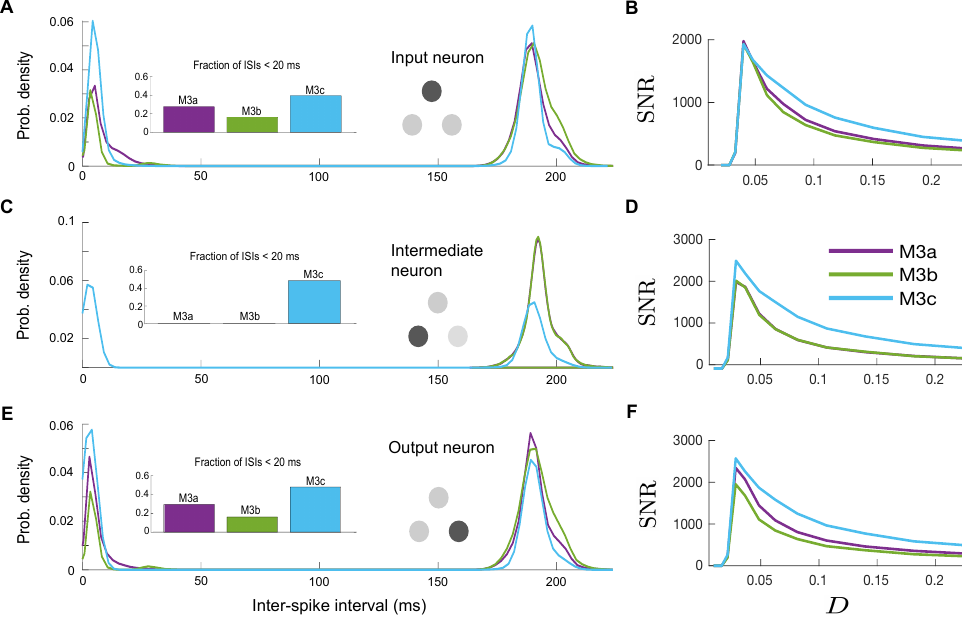}
\caption{Doublet formation and ISI per FFL neuron. (A), (C), and (E) show the empirical ISI probability densities and the fraction of short ISIs ($< 20$~ms; inset), while (B), (D), and (F) show the corresponding SNR curves for the input, intermediate, and output neurons, respectively. Colours: M3a (purple), M3b (green), and M3c (blue). Parameters: $p = 0.05$, $W_\text{intra} = 30$, $W_\text{inter} = 15$, and $I = 3.75$.}

\label{fig:doubletFFL}
\end{figure}

To further investigate this relationship, we analyze doublet formation and synchrony at the neuron level within the FFL motifs. Specifically, we examine how doublet activity varies across the input, intermediate, and output neurons by computing the ISI distributions and the fraction of short ISIs ($<20$~ms) for each neuron type in motifs M3a–c. Figs.~\ref{fig:doubletFFL}A, C, and E show the ISI probability densities for the input, intermediate, and output neurons, respectively, with the corresponding fraction of short ISIs displayed in the inset of each panel.

Across all neuron types, M3c exhibits the largest fraction of short ISIs, consistent with its recurrent connection between the input and intermediate neurons. M3a surpasses M3b despite lacking recurrence, indicating that the placement of feedback connections is more consequential than their presence. Figs.~\ref{fig:doubletFFL}B, D, and F show the corresponding SNR curves for the input, intermediate, and output neurons. The SNR profiles follow the same ordering observed in the ISI distributions: M3c attains the highest SNR across a broader range of noise intensities, while M3a and M3b display nearly coincident, lower-amplitude curves. This hierarchy, $\text{M3c}>\text{M3a}>\text{M3b}$, persists across all neuron types, linking the frequency of short ISIs to noise-induced coherence. These results demonstrate that motifs with more frequent doublets exhibit greater synchrony, but the relationship is nontrivial: M3b includes a recurrent link yet produces fewer doublets and a lower SNR than M3a, showing that synchrony depends on the specific placement of recurrent connectivity rather than its presence alone. Overall, M2 and M3c display the greatest fraction of doublet bursts, consistent with their role as motifs with a high propensity for noise-induced synchrony.

\subsection{Effect of network size}

\begin{figure}[t]
    \centering
    \includegraphics[width=\linewidth]{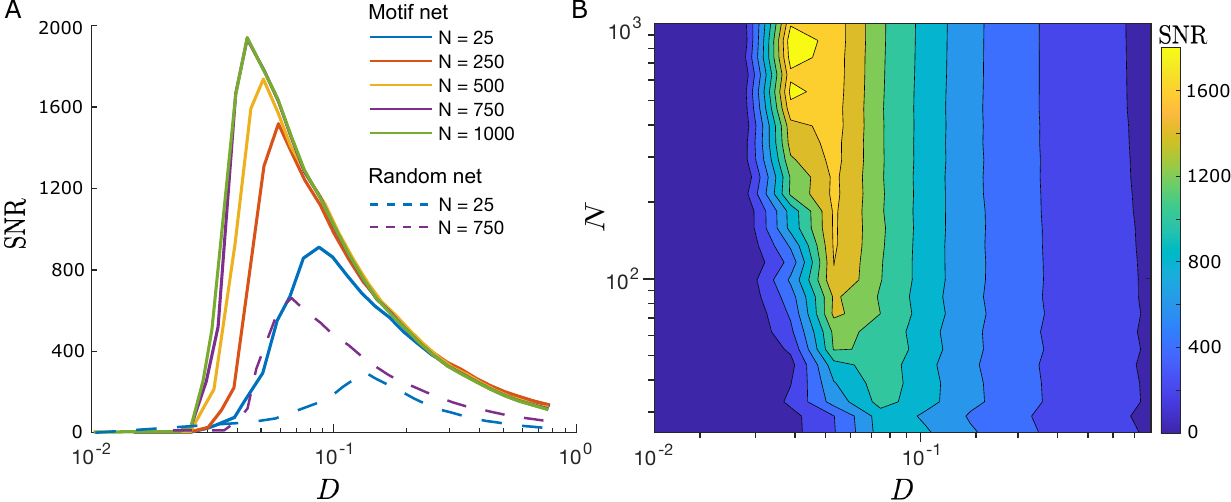}
    \caption{Impact of network size on synchronization. (A) SNR as a function of noise intensity $D$ for different network sizes $N$. Solid blue, orange, yellow, purple, and green curves correspond to motif networks with $N = 25$, $250$, $500$, $750$, and $1000$, respectively (the purple and green curves nearly coincide). Dashed blue and purple curves indicate random control networks for $N = 25$ and $N = 750$. (B) Heat map of SNR as a function of $D$ and $N$. Parameters: $I = 3.75$, $W_\text{intra} = 30$, $W_\text{inter} = 15$, and $p = 0.05$.}
    \label{fig:network size}
\end{figure}

Having characterized synchrony for a fixed network size ($N=1000$), we next examine its dependence on network size by evaluating the SNR as a function of $N$. The results are shown in Fig.~\ref{fig:network size}A, where solid curves correspond to motif networks of size $N = 25$ (blue), $250$ (orange), $500$ (yellow), $750$ (purple), and $1000$ (green). Increasing $N$ increases the SNR and shifts the optimal noise intensity $D^*$ to smaller values, indicating that larger networks require weaker stochastic input for maximal synchronization. For example, the $N=25$ network attains a peak SNR of $\approx 900$ at $D^* \approx 0.09$, whereas the $N=500$ network reaches $\approx 1700$ at $D^* \approx 0.06$. Beyond $N \approx 750$, the SNR curves coincide, indicating that coherence enhancement and noise sensitivity both saturate for larger networks.
The random control networks (dashed curves in Fig.~\ref{fig:network size}A) exhibit consistently lower SNR across all noise intensities. Even at $N=750$, the random network’s SNR remains below that of the motif network with $N=25$, indicating that structural organization, rather than size, primarily determines synchronization capacity. Moreover, the random networks require higher noise intensities to reach their maximal SNR, which remains substantially lower than that of the motif networks at comparable $N$.

The heat map in Fig.~\ref{fig:network size}B (SNR versus $D$ and $N$) corroborates these findings. For each $N$, a distinct SNR maximum is observed, consistent with CR across all network sizes. As $N$ increases, both peak SNR and $D^*$ evolve monotonically---the former increasing and the latter decreasing---until they converge when $N \approx 750$. Beyond this critical size, further growth yields negligible improvement, indicating a regime where synchronization dynamics become effectively size-invariant.

\subsection{Effects of network heterogeneity}
\label{sec:hetero}

\begin{figure}[t]
\centering
\includegraphics[width=1\linewidth]{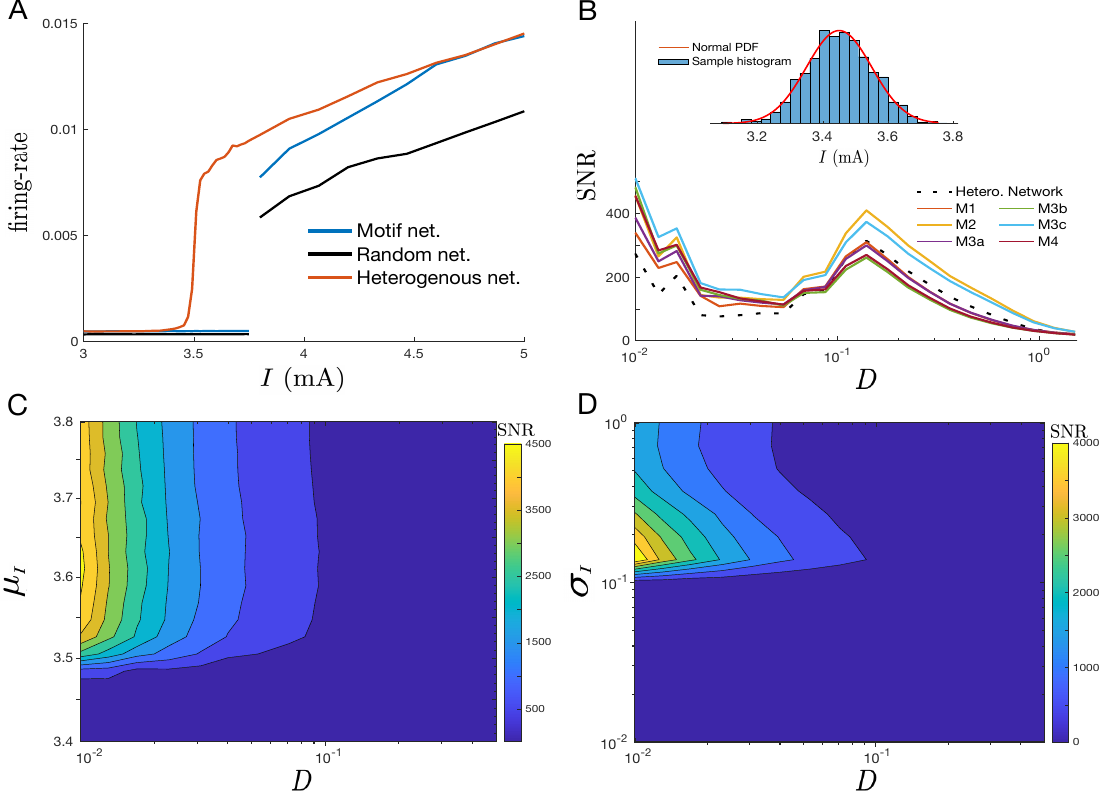}
\caption{Effects of network heterogeneity. (A) Firing-rate (of deterministic network) as a function of applied current $I$. The red, blue, and black curves represent the heterogeneous network ($\mu_I = 3.45$, $\sigma_I = 0.15$), the motif network ($p = 0.05$), and the random network ($\overline{p} = 0.0507$), respectively. (B) SNR as a function of $D$ per motif for a heterogeneous network with $\mu_I = 3.45$ and $\sigma_I = 0.15$. The inset shows the distribution of applied currents $I_i$ across neurons for the heterogeneous network. (C)–(D) Heat maps of SNR as functions of $\mu_I$ and $D$ (panel C, $\sigma_I = 0.15$) and $\sigma_I$ and $D$ (panel D, $\mu_I = 3.45$). Parameters: $W_\text{intra} = 30$ and $W_\text{inter} = 15$.}
\label{fig:hetero}
\end{figure}

In the preceding sections, we examined homogeneous networks of identical neurons initialized with random conditions. We now introduce heterogeneity by varying the applied current $I_i$ across neurons according to a normal distribution, $I_i \sim \mathcal{N}(\mu_I, \sigma_I^2)$, thereby introducing variability in intrinsic spiking thresholds. The inset of Fig.~\ref{fig:hetero}B shows a sample histogram of $I_i$ values drawn from $\mathcal{N}(3.45, 0.15^2)$, illustrating the distribution of applied currents within the heterogeneous network.

Fig.~\ref{fig:hetero}A shows the average spike rate as a function of applied current $I$ for the motif, random, and heterogeneous networks. Both the motif and random networks exhibit spiking profiles analogous to class II excitability \cite{izhikevich2000neural}, with an abrupt onset of activity near $I \approx 3.8$, consistent with the deterministic single-neuron model, which is class II excitable. In contrast, the heterogeneous network displays behaviour consistent with class I excitability, where the spike rate increases continuously from zero. This transition arises from the distributed applied currents $I_i$, which generate a mixed population of excited and excitable neurons, shifting the excitation threshold leftward to $I \approx 3.48$. Across all $I$, the heterogeneous network maintains a higher mean spike rate than the motif network, most notably near threshold, though the difference diminishes as $I$ increases. The elevated firing rate at low currents may partly reflect the mixed excitability of the population, with neurons near the upper tail of the current distribution contributing disproportionately to population-level spiking, allowing sustained higher activity even for $I > 3.8$.

To examine how heterogeneity affects coherence resonance, we set $\mu_I = 3.45$, slightly below the excitation threshold of the heterogeneous network. Fig.~\ref{fig:hetero}B shows the motif- and network-level SNR for this configuration with $\mu_I = 3.45$ and $\sigma_I = 0.15$. The network exhibits coherence resonance with an optimal noise intensity near $D \approx 0.15$, substantially higher than that of the motif network placed just above threshold. The corresponding peak SNR is $\approx 300$, markedly lower than in the motif network, indicating that despite the higher mean firing rate, the heterogeneous network displays less ordered spiking activity. At the motif level, M2 and M3c continue to yield the highest SNR, consistent with previous results, whereas M3b and M4 remain the least coherent. The SNR curves further show that weak noise initially reduces coherence before a modest increase near the resonance peak, after which higher noise intensities again suppress synchrony.

Figs.~\ref{fig:hetero}C–D summarize how the distribution of applied currents influences coherence and SNR. In Fig.~\ref{fig:hetero}C, we vary the mean current $\mu_I$ and noise intensity $D$ while fixing $\sigma_I = 0.15$. For $\mu_I > 3.5$, where most neurons are excited, the addition of noise monotonically reduces SNR. For $\mu_I < 3.5$, SNR remains low across all noise intensities, with $\text{SNR}<500$ for $\mu_I < 3.48$, consistent with Fig.~\ref{fig:hetero}B and lower than the motif network. In Fig.~\ref{fig:hetero}D, we examine the effect of input variance by varying $\sigma_I$ and $D$ while fixing $\mu_I = 3.45$, where heterogeneity is represented by the variance $\sigma_I^2$ of the applied current distribution. An optimal level of heterogeneity emerges around $\sigma_I \approx 0.15$ across noise intensities, beyond which coherence degrades and below which the SNR remains at a low level ($<500$). Since $\sigma_I$ is likely dependent on $\mu_I$, the precise location of this optimum likely depends on their joint distribution.

\section{Discussion and future work}
\subsection{Discussion}
\label{sec:discussion}
This study examined how local motif architecture shapes noise-induced synchronization after motifs are embedded in a larger neuronal network. Across the parameter regimes considered, both the full network and its constituent motif classes exhibit a non-monotonic dependence of coherence on noise intensity, consistent with CR. The magnitude of this response, however, depends strongly on connection arrangement. Relative to a random network matched in mean in-degree and mean synaptic strength, the motif-structured network attains greater coherence and reaches its optimum at weaker noise. Thus, the response cannot be characterized by connection number and average coupling strength alone; the paths formed by strong local connections also matter.

The motif-level results identify M2 and M3c as particularly coherent structures under the baseline conditions. This difference is not explained simply by motif size or recurrence. The three FFL variants contain the same number of neurons, yet their coherence differs, and M3a can exceed M3b despite lacking a recurrent connection. M3b and M3c both contain recurrence but place the recurrent edge at different positions. The resulting differences show that recurrence does not have a single dynamical effect: its influence depends on where it occurs within the local circuit. The rewiring analysis reaches the same conclusion at the network level. Preserving the number and strength of strong local synapses while disrupting the M2 or M3c arrangement reduces network coherence more strongly than rewiring several other motif classes, whereas rewiring some motifs has little effect or can slightly increase coherence. Strong local coupling is therefore not uniformly beneficial; its effect depends on how those connections are arranged.

The spike-timing analysis provides a possible dynamical correlate of these structural differences. M2 and M3c produce the largest fractions of short inter-spike intervals and frequent spike doublets, and the FFL neuron-level analysis shows that the motif with the most frequent short intervals also has the largest SNR across input, intermediate, and output neurons. This association does not establish that doublets cause greater coherence, but it shows that the motifs with the strongest noise-induced coherence also organize spike timing differently. The comparison between M3a and M3b again cautions against attributing this effect to recurrence alone: the non-recurrent M3a can produce more short intervals and greater coherence than M3b. The specific placement of feedback is therefore more informative than its mere presence.

Inter-motif connectivity and coupling strength regulate how strongly these local differences are expressed. Increasing the inter-motif connection probability produces a sharp rise in coherence in the sparse-connectivity regime, followed by diminishing gains. Increasing intra- or inter-motif coupling also enhances network coherence, but redistributing a fixed coupling budget between local and background connections affects motif classes differently. M2 and M3c benefit most as coupling becomes more intra-motif dominated, whereas other classes can depend more strongly on background input. Local and global coupling therefore play complementary roles: intra-motif connections organize activity within motifs, while inter-motif connections coordinate those local responses across the network.

The network-size analysis shows that the structural advantage of the motif network is not a small-network artifact over the range examined. Peak coherence increases and the optimal noise intensity decreases as $N$ grows, with both trends beginning to saturate near the largest sampled sizes. The matched random controls remain less coherent and require stronger noise. These results indicate that increasing size changes the magnitude and noise sensitivity of the response without removing the difference between structured and randomized local connectivity.

Applied-current heterogeneity weakens coherence but does not erase the relative motif differences. Distributing $I_i$ across neurons broadens the transition from quiescence to spiking, lowers the peak SNR, and shifts the coherence optimum toward stronger noise compared with the homogeneous network. Nevertheless, M2 and M3c remain among the most coherent motif classes. The persistence of this ordering suggests that the influence of local arrangement is robust to moderate variation in neuronal excitability, even though heterogeneity changes the overall operating regime of the network.

These conclusions are specific to the idealized architecture studied here. The motif proportions are imposed, inter-motif connectivity is Erd\H{o}s--R\'enyi, the noise is independent and additive, and the neuronal model is a reduced Izhikevich system. The rewiring experiments establish that arrangement matters once motifs are present, but they do not explain how such motifs arise or whether activity-dependent plasticity would generate or stabilize the same structures. Within these limitations, the central result is that strong local connections are not dynamically equivalent when arranged differently: motif organization can remain visible at the population level after embedding in a large stochastic network.

\subsection{Future work}
\label{sec:futurework}
A key direction for future research is to determine how network motifs contribute to the emergence of functional communities within biologically plausible neural circuits. In particular, it remains to be established which motifs form the structural and dynamical cores of such communities, and how their spatial and functional organization influences community detectability and structure.

\begin{figure}[t]
\centering
\includegraphics[width=0.85\linewidth]{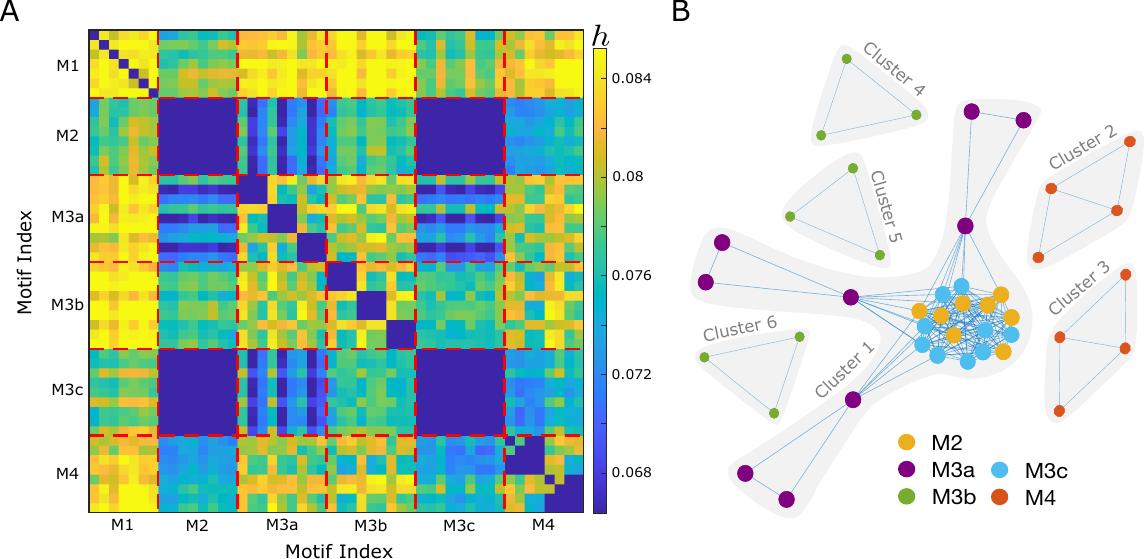}
\caption{Community detection. (A) Hamming distance ($h$) dissimilarity matrix for a 50-neuron network. Cooler colours correspond to smaller $h$ values, indicating greater similarity. (B) Clustered community network derived from the dissimilarity matrix in panel~(A). Parameters: $I = 3.75$, $p = 0.05$, $W_\text{intra} = 30$, and $W_\text{inter} = 15$.}
\label{fig:community}
\end{figure}

To explore this question, we performed a preliminary analysis based on spike-train similarity. Binary spike trains were compared using the Hamming distance $h$, following the method of \cite{humphries2011spike}, and pairwise dissimilarities were thresholded at the 10th percentile to generate functional connectivity graphs. The resulting community structure, shown in Fig.~\ref{fig:community}B, was derived from the 50-neuron dissimilarity matrix in Fig.~\ref{fig:community}A. Prominent clusters emerged around M2 and M3c motifs, suggesting that motifs with strong internal coherence and recurrent coupling may serve as local hubs of synchronization. In contrast, M1 neurons, which lack internal connectivity, did not form stable communities. Together, these findings indicate that recurrently coupled motifs may provide the structural basis for localized functional organization within the network.

Future work should address two main challenges. First, threshold- and $k$-means–based community detection methods introduce arbitrary parameters that may obscure underlying structure or report false positives; parameter-free approaches should therefore be employed to improve robustness. Second, incorporating time-resolved, asymmetric similarity measures could capture directional dependencies between spike trains and reveal the functional organization of motif-driven communities in higher resolution.

\appendix
\section{Bifurcation analysis of the deterministic model}
\label{appendix:bifurcation}
\begin{figure}[t]
    \centering
    \includegraphics[width=1\linewidth]{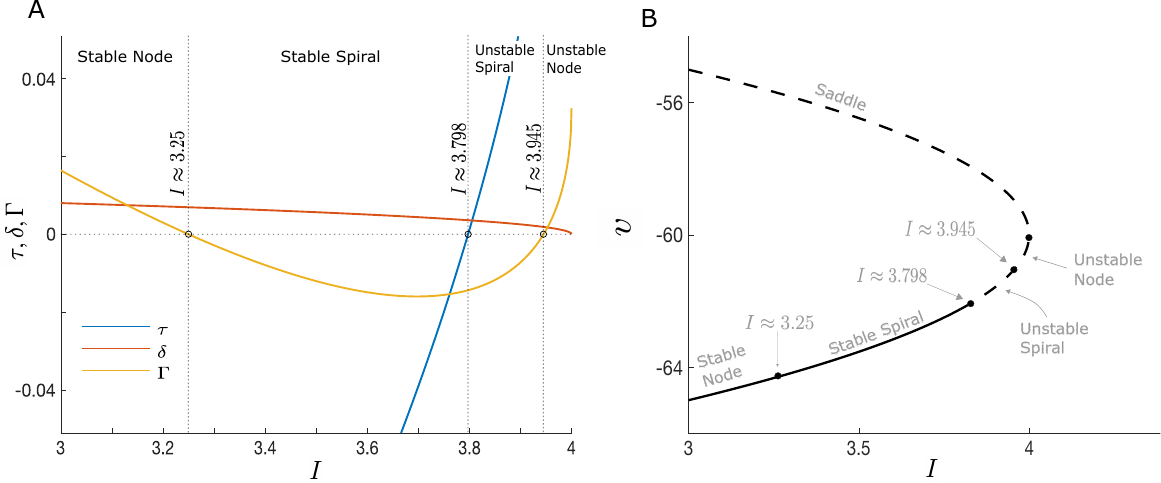}
   \caption{Appendix Fig.~A1. Trace ($\tau$), determinant ($\delta$), and discriminant ($\Gamma$) of the Jacobian at $(v_1^*,u_1^*)$ as functions of applied current $I$. Blue, red, and orange curves represent $\tau$, $\delta$, and $\Gamma$, respectively. The bifurcation diagram of $v$ versus $I$ is shown below; solid lines denote stable equilibria and dashed lines denote unstable equilibria.}
    \label{fig:tau-delta}
\end{figure}

We analyze the bifurcation structure of a single \textit{regular spiking} Izhikevich neuron in the deterministic regime using linear stability analysis. In the deterministic and uncoupled regime ($D = 0$, $W_\text{intra} = 0$, $W_\text{inter} = 0$), the model reduces to 
\begin{align} 
\frac{\mathrm{d}v}{\mathrm{d}t} &= 0.04v^2 + 5v + 140 - u + I, \label{appendix_v}\\
\frac{\mathrm{d}u}{\mathrm{d}t} &= a(bv - u). \label{appendix_u}
\end{align}
The system is in equilibrium if and only if $\mathrm{d}v/\mathrm{d}t = \mathrm{d}u/\mathrm{d}t = 0$. From Eq.~\eqref{appendix_u}, $\mathrm{d}u/\mathrm{d}t =0$ implies $u = bv$. Substituting this into Eq.~\eqref{appendix_v} yields the equilibrium condition 
\begin{equation*}
0.04v^2 + (5 - b)v + (140 + I) = 0.
\end{equation*}
Hence, the equilibria are 
\begin{equation*}
(v^*,u^*) = \left(12.5\left[b - 5 \pm \sqrt{(5 - b)^2 - 0.16(140 + I)}\right],\, b v^*\right).
\end{equation*}
The number of equilibria depends on the discriminant $\Delta = (5 - b)^2 - 0.16(140 + I)$, which vanishes at $I = 6.25(5 - b)^2 - 140$. To assess stability, we consider the Jacobian matrix
\begin{equation*}
J(v^*,u^*) = 
\begin{pmatrix}
0.08v^* + 5 & -1 \\
ab & -a
\end{pmatrix},
\end{equation*}
with trace, determinant, and discriminant given by
\begin{equation*}
\tau = 0.08v^* + 5 - a, \qquad
\delta = -a(0.08v^* + 5) + ab, \qquad
\Gamma = \tau^2 - 4\delta.
\end{equation*}
The dependence of $\tau$, $\delta$, and $\Gamma$ on $I$ is shown in Fig.~\ref{fig:tau-delta}A, which depicts the $\tau$–$\delta$–$\Gamma$ plane and identifies critical transitions in stability. Fixing $a = 0.02$ and $b = 0.2$, three distinct dynamical regimes emerge for $I < 4$, $I = 4$, and $I > 4$.

When $I < 4$, there are two equilibria,
\begin{align*}
(v_1^*,u_1^*) &= \left(12.5\left[b - 5 - \sqrt{(5 - b)^2 - 0.16(140 + I)}\right],\, b v_1^*\right), \\
(v_2^*,u_2^*) &= \left(12.5\left[b - 5 + \sqrt{(5 - b)^2 - 0.16(140 + I)}\right],\, b v_2^*\right).
\end{align*}
The first equilibrium $(v_1^*,u_1^*)$ is stable when $\tau < 0$ and $\delta > 0$, while $(v_2^*,u_2^*)$ is a saddle when $\delta < 0$. A transition in stability occurs when $\tau = 0$, corresponding to
\begin{equation*}
0.18 - \sqrt{4.8^2 - 0.16(140 + I)} = 0 \quad \Rightarrow \quad I \approx 3.798.
\end{equation*}
Thus,  $(v_1^*,u_1^*)$ is stable for $I < 3.798$ and unstable for $3.798 < I < 4$. The discriminant $\Gamma$ further distinguishes between stable nodes ($I < 3.25$) and stable spirals ($3.25 < I < 3.798$), as shown in the $\Gamma(I)$ curve in Fig.~\ref{fig:tau-delta}A.

When $I = 4$, the system has a single equilibrium at $(v^*,u^*) = (-60, -12)$, where $\delta = 0$, implying a zero eigenvalue and a non-isolated critical point. This corresponds to the fold bifurcation point identified in Fig.~\ref{fig:tau-delta}B,.

When $I > 4$, no equilibria exist, $v$ grows without bound and is reset by the model’s threshold mechanism. This regime corresponds to tonic spiking, where the spike frequency $f$ increases approximately linearly with the applied current $I$, as summarized in the bifurcation diagram in Fig.~\ref{fig:tau-delta}B.

Fig.~\ref{fig:tau-delta}B summarizes the results of the $\tau$–$\delta$–$\Gamma$ analysis and the equilibrium classification above, illustrating the transition from excitable to tonic-spiking behaviour near $I \approx 3.8$. For $I < 3.25$, the neuron exhibits a stable fixed point; for $3.25 < I < 3.798$, stable spirals and small oscillations arise; and for $I > 3.798$, the system transitions to tonic spiking.

\section*{Acknowledgments}
This work was supported by a Natural Sciences and Engineering Research Council of Canada (NSERC) grant to Na Yu and an Ontario Graduate Scholarship (OGS) to Gurpreet Jagdev, with additional support from Toronto Metropolitan University.

\bibliographystyle{unsrt}
\bibliography{references}

\end{document}